# Designing a Transition Photonic Band with a Synthetic Moiré Sphere

Z. N. Liu[1], X. Q. Zhao[1], J. Yao[2], C. Zhang[1], J. L. Xu[1], S. N. Zhu[1], and H. Liu[1,*]

[1]*National Laboratory of Solid State Microstructures, School of Physics, Collaborative Innovation Center of Advanced Microstructures, Nanjing University, Nanjing 210093, China*

[2]*Department of Materials Science and Engineering, University of California, Berkeley, California 94720, USA*

Email: liuhui@nju.edu.cn

In recent years, twisted bilayer graphene has become a hot topic and inspired the research upsurge of photonic moiré lattice. Here, we designed a photonic moiré superlattice with two synthetic twist angles and constructed a synthetic moiré sphere based on these two angles. Thus, we have more degrees of freedom to design the band structure flexibly. A type of transition photonic bands (TPBs) is obtained in such a moiré superlattice. We investigate the influence of two twist angles on TPBs and find a series of magic angle pairs with optimal band compression of TPB. The interesting optical properties of TPBs are experimentally demonstrated, including pulse delay, nonlinear optical enhancement, and pulse width compression. Our work introduces a new path to explore multi-twist angles moiré superlattices and reveals that the designed photonic moiré superlattice based on moiré spheres has broad application prospects including optical signal processing, nonlinear optics processes ,and other light-matter interactions.

***Introduction.*** —A moiré lattice is a composite structure formed by the overlap of two identical or similar periodic structures[1]. Recently, in twisted double-layer graphene, it was found that at the so-called magic angle, the band appears as a flat band near the Fermi level, and there is a nontrivial topological phase [2-4]. Moreover, the Mott insulating phase and superconducting phase are exhibited by double-layer graphene [5-8]. Meanwhile, the moiré lattice brings the possibility of exotic physics phenomena, including moiré excitons [9], fractional Chern insulators [10], lattices with competing periodicities [11], etc.

On the other hand, moiré lattices have also introduced interesting physical effects in photonic systems [1,12,13]. Especially, moiré fringes have been applied to metal surfaces to modulate plasmonic dispersion and group velocity [13]. In a two-dimensional photonic moiré lattice, the localization-delocalization transition of light is realized experimentally [14]. The tunable topological transitions and phonon polaritons were achieved in bilayers of α-phase molybdenum trioxide (α-MoO3) [15,16]. The formation of optical solitons are controlled by the twist angle in photonic moiré lattices [17,18] and magic-angle lasers in nanostructured moiré lattice exhibit salient features [19]. Recently, a coupled-mode theory for low-angle twisted bilayer honeycomb photonic lattices reveals a correspondence between fermionic and bosonic moiré systems [20]. And the so-called magic distance was found in the bilayer photonic moiré lattice. Meanwhile, slow light, nonlinear effects, chiral plasmons, thermal emitters, and filters exhibit excellent properties in the photonic moiré lattice system [21,22]. One major attraction of these structures is that their optical properties are strongly dependent on the twist angle. However, up to now, in most reported works, the bands of moiré lattices are tuned by changing a single twist angle. The degree of freedom of a single twist angle is very limited. In most moiré structures, it is difficult to find interesting physical properties by adjusting a single twist angle. Fortunately, synthetic parameters allow us to explore physical phenomena in high-dimensional synthetic space [23-33]. Inspired by this method, we will introduce synthetic twist angles are introduced into the moiré system.

In this work, we propose a moiré superlattice with two synthetic twist angles in which a kind of TPBs is obtained. A two-dimensional synthetic moiré sphere is constructed based on these two synthetic twist angles. The twist angles are tuned to design the moiré lattice. A series of magic angle pairs on the moiré sphere are found to obtain TPBs with optimal band compression. Experimentally, we directly investigate the properties of TPBs, including narrow-band filtering, pulse delays, nonlinear optical enhancement, and pulse width compression.

***Design of photonic moiré superlattice based on Synthetic moiré sphere.*** —Firstly, let us consider a simple photonic lattice, which has two refractive index layers of equal thickness, $n_a$ and $n_b$, in a unit cell, whose refractive index distribution can be characterized as

$$V_\Lambda(z) = \sin(\frac{2\pi}{\Lambda}z + \frac{\pi}{2}), \tag{1}$$

, where $\Lambda$ is the spatial period and z is the spatial location. When the spatial position is $V_\Lambda(z) > 0$, the refractive index distribution function $n_\Lambda(z) = n_a$, otherwise, $n_\Lambda(z) = n_b$. To illustrate the idea of the moiré superlattice, we consider a 1D superlattice composed of the two simple photonic lattices with different periods, $\Lambda_1$ and $\Lambda_2$. The moiré superlattice can be expressed as

$$P_\Lambda(z) = \sin^2\frac{\gamma}{2} \cdot V_{\Lambda_1}(z) + \cos^2\frac{\gamma}{2} \cdot V_{\Lambda_2}(z), \tag{2}$$

, where $\gamma \in (0, 180°)$, and $V_{\Lambda_1}(z)$, $V_{\Lambda_2}(z)$ is the simple lattices structure function and $P_\Lambda(z)$ is the moiré superlattice structure function. And when $P_\Lambda(z) > 0$, the refractive index equals $n_a$; otherwise, $n_b$. In general, $\Lambda_1$ and $\Lambda_2$ vary continuously. When $\Lambda_1$ and $\Lambda_2$ are incommensurable, the moiré superlattice is not periodic and the band cannot be obtained by Bloch's theorem. Therefore, we consider the case where $\Lambda_1$ and $\Lambda_2$ are commensurable, i.e. $\Lambda_1/\Lambda_2 = m/n$, where m, n are positive integers. Then, the period of the superlattice is $\Lambda = n\Lambda_1 = m\Lambda_2$. If $\Lambda_1$ and $\Lambda_2$ are changed and the ratio $\Lambda_1/\Lambda_2$ keeps unchanged, the bands only shift in wavelength, but the band configuration does not change. Therefore, the band configuration is mainly determined by the ratio m/n. Here, we can define the synthetic twist angle α=8×arctan(m/n), which plays an important role in tuning the band configuration. In Fig. 1(i), α is shown as a synthetic angle in the parameter space $(\Lambda_1, \Lambda_2)$ visually. The line passing through the lattice point, whose angle with $\Lambda_2$ axis α/8 represents a periodic moiré superlattice. Otherwise, the line with angle α/8 that doesn't pass through any lattice point represents a non-periodic moiré superlattice. In Fig. 1(j), γ can also be represented as a synthetic angle in the parameter space $(V_{\Lambda_1}(z), V_{\Lambda_2}(z))$. The structure functions of two simple photonic lattices as basis vectors $(V_{\Lambda_1}(z), V_{\Lambda_2}(z))$, construct the synthetic parameter space. And γ/2 can be represented as the spatial twist angle in the synthetic parameter space, marked by red letters in Fig. 1(j). $\sin^2\frac{\gamma}{2} \cdot V_{\Lambda_1}(z)$ and $\cos^2\frac{\gamma}{2} \cdot V_{\Lambda_2}(z)$ can be obtained by twice triangular transformations of the two basis vectors $V_{\Lambda_1}(z)$, $V_{\Lambda_2}(z)$ to the line with the angle γ/2 with $V_{\Lambda_2}(z)$ axis. The projection of the basis vector $V_{\Lambda_2}(z)$ to the line at angle γ/2 becomes $\cos\frac{\gamma}{2} \cdot V_{\Lambda_2}(z)$ and then to the basis vector $V_{\Lambda_2}(z)$ becomes $\cos^2\frac{\gamma}{2} \cdot V_{\Lambda_2}(z)$. Meanwhile, the sine of the basis vector $V_{\Lambda_2}(z)$ to the line at angle γ becomes $\sin\frac{\gamma}{2} \cdot V_{\Lambda_1}(z)$ and then to the basis vector $V_{\Lambda_1}(z)$ becomes $\sin^2\frac{\gamma}{2} \cdot V_{\Lambda_1}(z)$. Therefore, γ/2 represents the proportion of the combined moiré superlattice occupied by the two initial lattice structures, $V_{\Lambda_1}(z)$ and $V_{\Lambda_2}(z)$. And when γ=0 or γ=180°, the moiré superlattice changes back to the original simple periodic photonic lattices structure. In the method, we design six superlattices as examples with two synthetic angles. For α=295.2° and 324.8°, the $V_{\Lambda_1}(z)$ and $V_{\Lambda_2}(z)$ are given as the red and blue lines in Fig. 1 (a, e), respectively. For γ=80°, 90°, and 100°, $P_\Lambda(z)$ are represented as red lines in Fig. 1(b, c, d, f, g, h). In fact, the moiré superlattices structures are well defined by two synthetic angles (α, γ). Therefore, we can construct a two-dimensional synthetic parameter space based on these two angles. In Fig. 1(k), there are two degrees of freedom on the sphere, the angle α with the x-axis in the horizontal plane, and the angle γ with the z-axis. In this way, all the structures of superlattices can be well defined by the points on the sphere. Here, we can call this parameter space a synthetic moiré sphere.

***Transition Photonic Band of moiré superlattice***—To analyze the effect of the synthetic twist angles on the bands, the bands of the moiré superlattices are calculated with the transfer matrix. A superlattice is taken as instance with α=237.6°=8×arctan(4/7) and $\Lambda = 7\Lambda_1 = 4\Lambda_2$. Its band structure is investigated with 0 < γ < 180° in the Brillouin zone $\left(0, \frac{2\pi}{\Lambda}\right)$. For γ = 0 and 180°, the superlattices are periodic lattice with $\Lambda_1$ and $\Lambda_2$, whose bands are given in Fig.2 (a) and (c) and show band foldings at boundary of Brillouin zone. In Fig.2(b), the change of band structures with γ is given as the band projection, the degeneracy introduced by the band folding is lifted, and a band gap emerges at two boundaries of Brillouin zone. The three bands (marked in green) above the first gap (marked in pink in Fig.2 (a)) on the left boundary (γ=0) are shifted to the bands below the first gap (marked in pink in Fig.2 (c)) on the right boundary (γ=180°). Here, we call the green bands as transition photonic bands (TPBs). For comparison, we show the band projections of two superlattices with α=284°=8×arctan(5/7) (m = 5, n = 7) and α=324.8°=8×arctan(6/7) (m = 6, n = 7) in Fig. 2 (d, e). There are two TPBs in Fig. 2(d), and one TPB in Fig. 2(e). Obviously, the number of TPBs is determined by |n-m|, which is understood from the band folding at two boundaries. It is worth stating that for the case of |n-m|>1, several TPBs are very close to each other and the gaps between TPBs is very small. In practical system, they are very easily merged together if the sample length is not large enough. However, for the case of |n-m|=1, we can obtain a single isolated TPB with narrow frequency intervals even for sample with only a few periods. Therefore, the isolated TPB is more useful for practical applications. In the following, we will only focus on the structures of |n-m|=1.

Tuning photonic bandwidth is an important means of manipulating light. For example, application often require narrow-band filtering in transmission and band flattening due to band compression achieves slow light and affects group velocity dispersion. Therefore, the tunability of the TPB bandwidth can be applied to manipulating light. To investigate the dependence of TPB bandwidth on γ, we define compression coefficient $g = \frac{band(\gamma = 180°)}{band(\gamma)}$, where *band*(γ) is TPB bandwidth with γ.

Obviously, the larger $g$, the narrower the TPB bandwidth. The varying bands of superlattice ($\alpha$=324.8°) with $\gamma$ are given in Fig. 3 (a) and the projection of these bands on the plane ($\omega$, $\gamma$) is given in Fig 3(b). The TPB is marked as green in both Fig. 3 (a) and (b). The optimal bandwidth compression of TPB can be found by tuning $\gamma$. As shown in Fig. 3(c), when $\gamma$=86.4°, $g$ reaches the maximum and the TPB bandwidth gets its minimum. In Fig. 3(d), the TPB at $\gamma$=86.4° (blue dots) is compared with the TPBs at $\gamma$=0 (black curve) and 180° (red curve), which has the narrower bandwidth than the other two cases. Beside the TPB, the several bands near TPB will also be compressed(see Supplemental Material, Sec. 1[34]). It means TPB is different from the resonance defect mode.

In the above discussion, we find that $g$ is determined by both $\alpha$ and $\gamma$ simultaneously. Therefore, we investigate the dependence of $g$ on the synthetic moiré sphere ($\alpha$, $\gamma$). For each $\alpha_m$, we can find the optimal compression coefficient $g_m$ at certain $\gamma_m$. Here, the corresponding found angle pairs ($\alpha_m$, $\gamma_m$) can be defined as magic angle pairs, which are shown in Table. Ⅰ. In Fig. 3(e), we plot the change $g_m$ with series magic angle pairs. As $\alpha_m$ approaches 360°, the $g_m$ is increased to infinity. However, if $\alpha$ equals 360°, the two periods are equal $(\Lambda_1 = \Lambda_2)$. Then the superlattices will change back to a simple periodic lattice, and TPBs disappears. Moreover, to show the distribution of magic angle pairs, we project the 3D spatial curve onto the 2D plane (see Fig 3(f)). It is obvious seen that ($\alpha_m$, $\gamma_m$) are located at two asymptotic curves. When m is odd, $\gamma_m$ approaches 90° along the red curve, and when m is even, $\gamma_m$ approaches 90° along the blue curve. Besides, we also investigate the dependence of TPBs on the incidence angle (Supplemental Material, Sec. 2[34]). For the TE mode, the results show the bandwidth of TPB decrease with $k_y$ increasing. And for the TM mode, the results show the bandwidth of TPB widens with $k_y$ increasing. This causes the group indices of the two mode to change in opposite directions(Supplemental Material, Sec. 3[34]).

| m | 1 | 2 | 3 | 4 | 5 | 6 | 7 |
|---|---|---|---|---|---|---|---|
| $g_m$ | 1.02 | 1.10 | 1.16 | 1.31 | 1.45 | 1.69 | 1.95 |
| ($\alpha_m$, $\gamma_m$) | (212.8°,111.6°) | (269.6°,89.8°) | (295.2°,93.6°) | (309.6°,87.2°) | (318.4°,91.8°) | (324.8°,86.4°) | (329.6°,91.0°) |
| m | 8 | 9 | 10 | 11 | 12 | 13 | 14 |
| $g_m$ | 2.35 | 2.82 | 3.49 | 4.30 | 5.44 | 6.84 | 8.83 |

| ($\alpha_m$, $\gamma_m$) | (332.8°, 87.2°) | (336.0°, 90.8°) | (338.4°, 87.4°) | (340.0°, 90.6°) | (341.6°, 87.6°) | (343.2°, 90.4°) | (344.0°, 87.8°) |
| --- | --- | --- | --- | --- | --- | --- | --- |

Table. I. series magic angle pairs ($\alpha_m$, $\gamma_m$), which is the corresponding values of α and γ for $g_m$.

*Experimental observation of transition photonic bands* —We fabricated three moiré superlattices: A (212.8°, 111.6°), B (295.2°, 93.6°), C (324.8°, 86.4°), which are marked as the dots in Fig. 1(i). Fig. 4(a) shows the SEM microscopy picture of the sample C. The TPB bandwidths of the three samples are obtained as 479nm, 117nm and 31nm. With α increasing, its transmission bandwidth is reduced (see Supplemental Material, Sec. 2[34]), which agree with our calculations. Through tuning the synthetic angles, we can flexibly design the TPB, which can be applied to realize filters with various bandwidths. In particular, narrow-band optical filters can be achieved by magic angle pairs (the synthetic angles modulating isolated TPB). And its transmittance can be designed by thickness, and the surface flatness is further improved. However, we are more concerned about phenomena such as pulse delay, nonlinear effects, and pulse width compression due to band compression.

As the TPB bandwidth is reduced, the group velocity can be reduced, which causes pulse delay. The change of pulse delay passing through the sample compared with the substrate is $\Delta \tau = \frac{L_\alpha}{v_g} - \frac{L_\alpha}{c} = \frac{L_\alpha}{c}(n_g - 1)$, where $L_\alpha$ is the thickness of the sample. If the influence of different $L_\alpha$ is ignored, the normalized pulse delay is $\Delta \tau_n = (n_g - 1)$, where $n_g$ is the group index. In our experiment, a homemade optical setup is established to measure the pulse delay of laser through multiphoton absorption by silicon chip. The experimental setup for the pulse delay measurement is shown in Fig. 5. It is based on a modified setup for measuring the relaxation time of carriers on silicon chips, which can generate time delays by shifting the optical path of the pump laser. However, in this work, we only focus on the jump in the reflection intensity of a silicon chip under the optical path change of the pump pulse laser. We use a laser at 800nm, which can be adjusted continuously as needed, 120 fs pulse duration, and 80 MHz repetition rate, and split a laser beam into two beams: one is the pump laser, and the other is the probe laser. The pump laser passes through the BBO crystal, whose wavelength becomes 400nm, and then passes through the chopper, where the repetition rate is modulated to amplify the signal by the lock-in amplifier. After that, the pump laser passes through the motorized stage, which can continuously adjust the optical path of the pump laser, and then the pump laser illuminates the silicon chip with the probe laser. Meanwhile, the probe laser passes through the sample (moiré superlattices or substrate), of which we need to measure pulse delay. The combined beam of pump laser and probe laser illuminates the silicon chip, and then the reflectivity intensity of the probe laser will be measured by the photodetector. Obviously, when the probe pulse illuminates the silicon chip earlier than the pump pulse, the reflection intensity is weaker due to the absorption of light by unexcited carriers in the silicon chip. As the motorized stage moves, the optical path difference between the pump laser and the probe laser decreases. When the probe laser illuminates the silicon chip later than the pump laser, the reflection intensity becomes stronger than before because the silicon chip carriers have been excited by the pump laser. Therefore, we can observe the jump line of the reflection intensity in Fig. 4(b)-(d). The time point of the jump is the position of the same optical path of the two beams, so the pulse delay time is obtained through the motorized stage. In Fig. 4(b)-(d), the red line is the spectrum of the silicon chip with the substrate placed on the optical path of the probe laser, and the blue line is the spectrum of the silicon chip with the moiré superlattice placed on the optical path of the probe laser. These blue lines move to the left due to the presence of the moiré superlattice.

When we place different samples on the light path of the probe laser, the optical path of the probe laser will change, and the position of the jump point will change. Therefore, we can obtain the pulse time delay according to the displacement of the jump point. For the setup error, the minimum measurement accuracy of the motorized stage is $\delta_l = 1um$. Therefore, in the whole process, the pulse delay theoretical error is $\delta_{delay} = \dfrac{2\delta_l}{c} \approx 0.01 ps$, where c is the speed of light in a vacuum.

In Fig.4(b), the difference between the two lines is 0.12ps, which is the measured pulse delay difference between the sample C and substrate. In this way, we can also measure the pulse delay of other two samples in Fig.4(c) and (d). In Fig. 4(e), the experimental data agree well with the calculated $\Delta\tau_n$ results. It can be seen that the increase of $g$ results in the increase of pulse delays. Meanwhile, we use the pulse delay measurement results to obtain the group index. As shown in Fig. 4(f), the red, green, and blue lines correspond to the group indices on different TPBs of moiré superlattices, and the red, green, and blue dots are experimental data, which were obtained according to pulse delay. Slow light with large group index and lower group velocity has potential applications in optical buffering and advanced time-domain optical signal processing. And slow light compresses optical energy in space, which enhance linear and nonlinear effects and so miniaturize functional photonic devices.

The band compression also leads to field enhancement, which enhances nonlinear optical effect. In Fig.4(g), we calculated the transmission electric field distribution at a wavelength on the TPB of different moiré superlattices. The electric field of the moiré superlattice with (324.8°,86.4°) is stronger near 800 nm. This enables the enhancement of the nonlinear effect. In experiments, we input an 800 nm laser into the three samples, and the produced second-harmonic generation (SHG) are measured, respectively in Fig. 4(h). Under the same incident laser intensity, the SHG generated by the sample C is enhanced nearly over 50 times stronger than the substrate. However, for the sample B, since the frequency of SHG is inside the bandgap, the SHG is weaker than the substrate. The experimental results can be well explained by the calculated results of field enhancement on the TPB. This also suggests that slow light can enhance light-matter interactions, such as nonlinearities on the TPB and the moiré superlattices enables the multiplication of the laser frequency to produce the SHG.

Meanwhile, we also measure the pulse shape of laser and the results show that due to stronger negative group velocity dispersion, pulse width is compressed by the TPB. We also analyzed the variation of the pulses of laser passing through the moiré superlattice. Variation in pulse width is affected by group velocity dispersion on the TPBs of the moiré superlattices. We have the pulse width variation formula,

$$\Delta\tau_p = D(\omega)L_\alpha\Delta\omega, \qquad (3)$$

where $D(\omega) = \dfrac{d(\dfrac{1}{v_g})}{d\omega}$ is the group velocity dispersion, $v_g$ is the group velocity, and $\Delta\tau_p$ is the pulse time width variation, $L_\alpha$ is thickness of the sample, and $\Delta\omega$ is the pulse spectral width. We define the pulse width rate as $\eta = \dfrac{\tau_0 + \Delta\tau_p}{\tau_0}$, where $\tau_0$ is the initial laser pulse width. In Fig. 6(a), we calculate the group velocity dispersion on TPBs for several moiré superlattice samples. Furthermore, according to the Eq. (3), we calculated the pulse width rate $\eta$ and compared it with the experimental results, which were measured by autocorrelator in Fig. 6(b). In a certain wavelength range, the pulse width with a large α changes more drastically. In Fig. 6(c), comparing the pulse after passing through the moiré superlattice with through the substrate, we demonstrate the moiré superlattice has the effect of compressing the pulse width. At the wavelength of 810 nm, sample C produces more pulse width compression than the other two samples, which is due to the larger negative group velocity dispersion at 810 nm. The properties of the moiré superlattice can also be used to fabricate a compact compressor for laser pulses. The pulse compressor allows the optical pulse to transform or maintain an ultrashort optical pulse during propagation, which will benefit ultrafast optical communications, optical signal processing, and non-communication application.

*Conclusion and discussion* —In summary, we propose a method of constructing photonic moiré superlattices with two synthetic twist angles. A two-dimensional synthetic moiré sphere is constructed based on these two angles, which gives all possible structural parameters of the moiré superlattice. We found that a series of magic angle pairs, corresponding to TPBs with optimal band compression. Meanwhile, the one-dimensional moiré superlattices is of advantages of convenient fabrication and low costs. Experimentally, a method to measure the pulse delay caused by the slow-light effect on the TPB of moiré superlattices by measuring the transient reflection spectrum on a silicon chip is proposed, and the enhancement of the SHG in the TPB is analyzed. The research results show band compression brings many interesting phenomena, such as pulse delay, pulse width compression, and nonlinear effects. High-dimensional, moiré superlattices with synthetic multi-twist angles exhibit many physical properties and physical phenomena. We believe that the designing method of synthetic multi-twist angles can be extended to other systems and many other synthetic moiré superlattices will be realized. Some special other interesting applications are possibly to be obtained based on revealing their peculiar optical properties.

Acknowledgements

This work was financially supported by the National Natural Science Foundation of China (Nos. 92163216 and 92150302).

# Figure

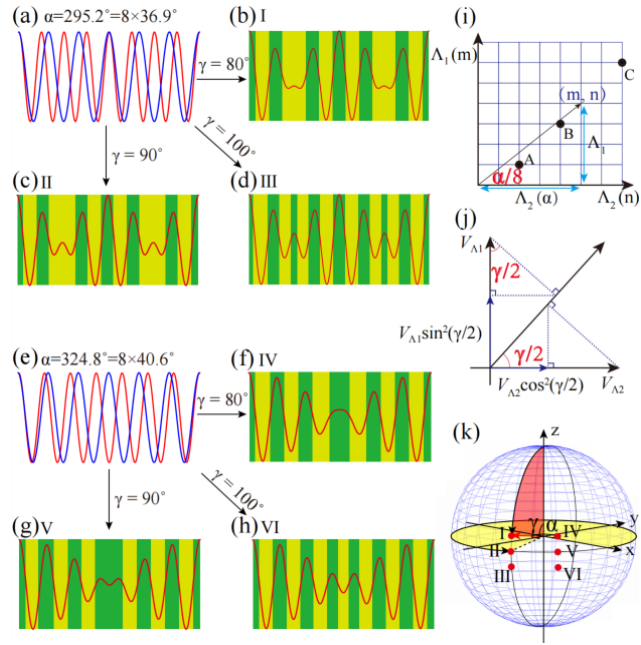

**Figure 1** (a) (e) The red and blue lines characterize two simple lattices, respectively. (a) The period ratio of the red line to the blue line is 3/4. (e) The period ratio of the red line to the blue line is 6/7. (b) (c) (d) (f) (g) (h) corresponds to structure: Ⅰ(295.2°,80°), Ⅱ(295.2°,90°), Ⅲ(295.2°,100°), Ⅳ(324.8°,80°), Ⅴ(324.8°,90°), Ⅵ(324.8°, 100°). The red lines are the characterization function and the background is schematic of the refractive index distribution. (i) The synthetic angle α in the parameter space ($\Lambda_1$, $\Lambda_2$). The black dots are the parameters of the experimental sample. (j) The synthetic angle γ in the parameter space $(V_{\Lambda_1}(z), V_{\Lambda_2}(z))$. (k) Synthetic moiré sphere defined with two synthetic angles (α, γ) and the superlattice defined with synthetic angles (red curve): I (295.2°,80°), II (295.2°,90°), III(295.2°,100°), IV(324.8°,80°), V(324.8°,90°), VI(324.8°, 100°).

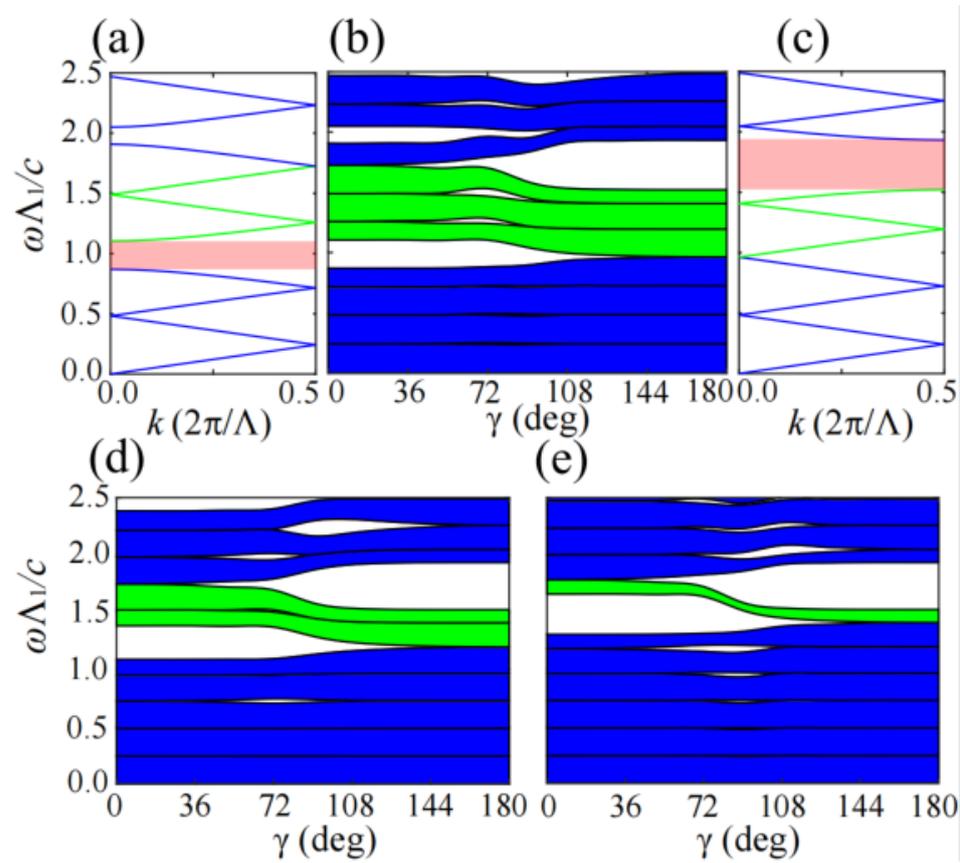

**Figure 2** The bands of simple lattices with (a) $\Lambda_1=\Lambda/7$, γ=180° and $\Lambda_2=\Lambda/4$, γ=180° in the Brillouin zone (0,2π/Λ). The pink shaded area is the first band gap, and the green line is TPB. The dependence of band projections of superlattice with (b) α=237.6°, (d)284°, (e) 324.8° on γ, respectively.

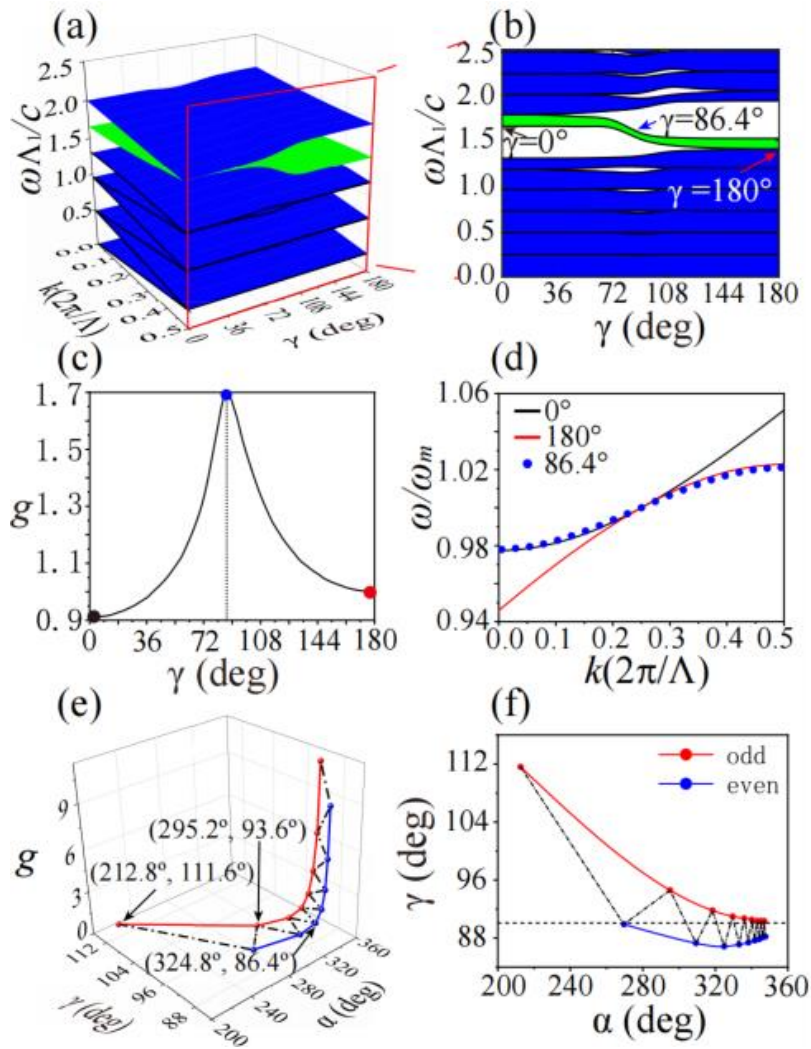

**Figure 3** (a) The bands of superlattice with α=324.8° and 0<γ<180°, and the green surface is the TPB. (b) is the band projection on the plane (ω, γ). (c) The change $g$ on γ. (d) Three bands of γ = 0, 86.4° and 180° and $\omega_m$ is midband positions. (e) The series $g_m$ with the magic angle pairs ($\alpha_m$, $\gamma_m$). (f) The two asymptotic curves of magic angles is when m is odd (red curve) and m is even (blue cure), respectively.

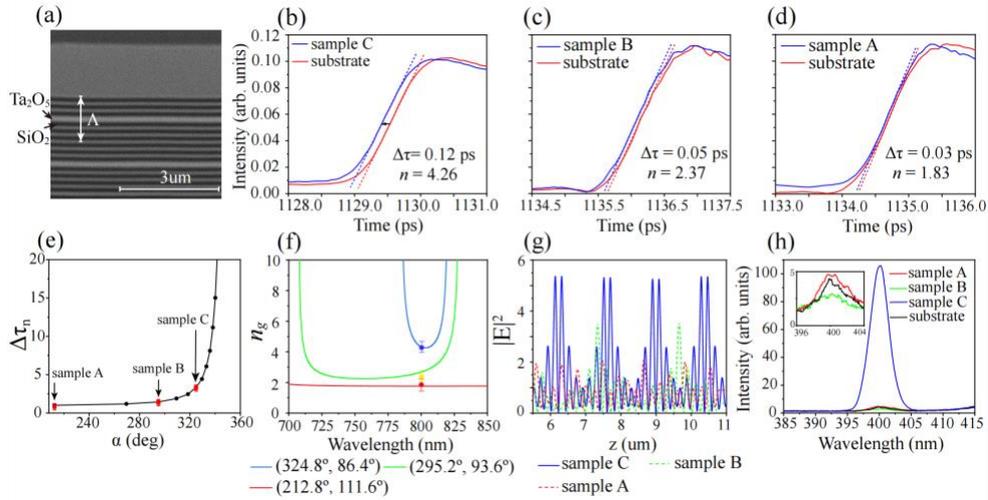

**Figure 4** (a) SEM microscopy picture of the sample C. (b)-(d) pulse time delay measurement results. The red line is the transient reflection intensity of the silicon chip with the probe laser passing through the substrate. The blue line is the transient reflection intensity of the silicon chip with the probe laser passing through the moiré superlattices. (b) corresponds to the sample C. (c) corresponds to the sample B. (d) corresponds to the sample A. The laser wavelength is 800nm. (e) Pulse delay of different moiré superlattices, the black dots are calculated data and the red squares are the above measured data of three samples. (f) The blue line is the group index on the TPB of the moiré superlattice with two synthetic angles ($\alpha$, $\gamma$): (324.8°,86.4°), the green line is the group index on the TPB of the moiré superlattice with (295.2°,93.6°), and the red line is the group index on the TPB of the moiré superlattice with (212.8°,111.6°). The red, green and blue dots are experimental results. (g) Electric field distributions of the sample A, B, C at a wavelength of 800 nm. (h) SHG intensity of the samples and substrate.

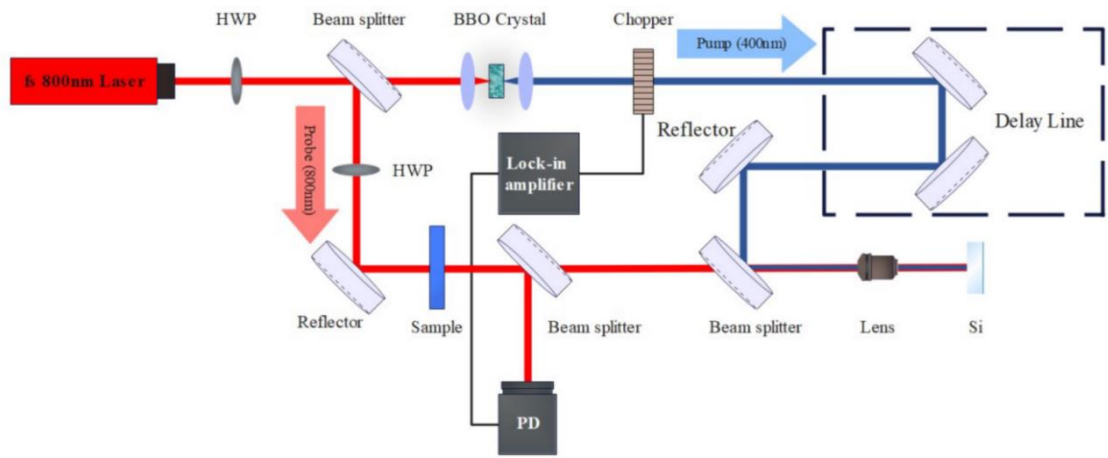

**Figure 5** Schematic diagram of the experimental setup for measuring pulse time delay.

The laser in the experiment is split into two beams through beam splitter. The one beam is called pump laser (blue rays) after passing through BBO crystal which can multiplier the laser, and the other beam is called probe laser (red rays). The pump laser can be delayed continuously by the delay line which can be moved precisely, then the pump laser and probe laser are focused together on the Si chip by the objective lens, finally, the reflected signal of the probe laser is collected by the photodetector. HWP, half wave plate; PD, photodetector.

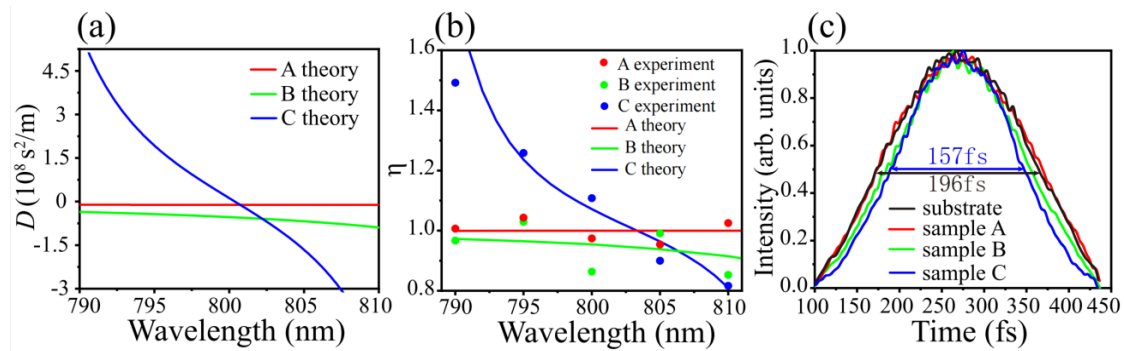

**Figure 6** (a) blue line is the group velocity dispersion on the TPB of the moiré superlattice with two synthetic angles (α, γ): (324.8°,86.4°), the green line is the group velocity dispersion on the TPB of the moiré superlattice with (295.2°,93.6°), and the red line is the group velocity dispersion on the TPB of the moiré superlattice with (212.8°,111.6°). (b) The pulse width rate before and after the laser passes through the moiré superlattices on the TPB. The solid line is the theoretical calculation results, the dot is the experimental results (c) shows laser pulse profiles. The black line is the original the pulse profile through the substrate, the red line is the original the pulse profile through the the sample A, the green line is the pulse profile through the sample B, and the blue line is the pulse profile through the sample C. The laser wavelength is 810nm.

# Supplementary Material for Designing a Transition Photonic Band with a Synthetic Moiré Sphere

Z. N. Liu[1], X. Q. Zhao[1], J. Yao[2] C. Zhang[1], J. L. Xu[1], S. N. Zhu[1], H. Liu[1,*]

[1]National Laboratory of Solid State Microstructures, School of Physics, Collaborative Innovation Center of Advanced Microstructures, Nanjing University, Nanjing 210093, China

[2]Department of Materials Science and Engineering, University of California,

Berkeley, California 94720, USA

**Section 1. The bands evolution analysis near the TPB**

In Fig. 1S(a), we show the dependence of the bandwidth of the fifth band, which is away from the TPB in Fig. 3(b), on $\gamma$. With the change of the moiré superlattice structure, the phenomenon of band narrowing also occurs in the other bands. And, we can explain this phenomenon of increasing band gap width and thus the band narrowing.

Firstly, we can define the concentration factor of the electric mode: $\sigma = \frac{\int_{\varepsilon_A} d^3z\, \varepsilon(z)\, |E(z)|^2}{\int d^3z\, \varepsilon(z)\, |E(z)|^2}$, which is an appropriate measure of the degree of concentration of the electric fields in the high-ε regions. Since the difference in the energy distribution of the band edge mode field above and below the band gap is responsible for the increase in the band gap, we can analyze the band gap increase due to moiré superlattice by calculating the concentration factor of the electric mode around the band gap.

We investigated the difference in the electric field energy distribution near the two band gaps above and below the fifth band in Fig. 3(b) with $\gamma$. In Fig. 1S(b), the red line represents the difference between the concentration factors of the two split modes at the edge of Brillouin zone below the fifth band: $\Delta\sigma_{5b} = \sigma_{5bb} - \sigma_{5ba}$, where $\sigma_{5bb}$ is the concentration factor of the electric mode below the band gap and $\sigma_{5ba}$ is the concentration factor of the electric mode above the band gap. Similarly, the blue line represents the difference between the concentration factors of the two split modes at the edge of the Brillouin zone above the fifth band: $\Delta\sigma_{5a} = \sigma_{5ab} - \sigma_{5aa}$, where $\sigma_{5ab}$ is the concentration factor of the electric mode below the band gap and $\sigma_{5aa}$ is the concentration factor of the electric mode above the band gap. For the fifth band, the increase of $\Delta\sigma_{5b}$ and $\Delta\sigma_{5a}$ represents the widening of its upper and lower band gaps, and the fifth band is compressed. When $\Delta\sigma_{5b} + \Delta\sigma_{5a}$ is maximum, the band bandwidth is narrowest.

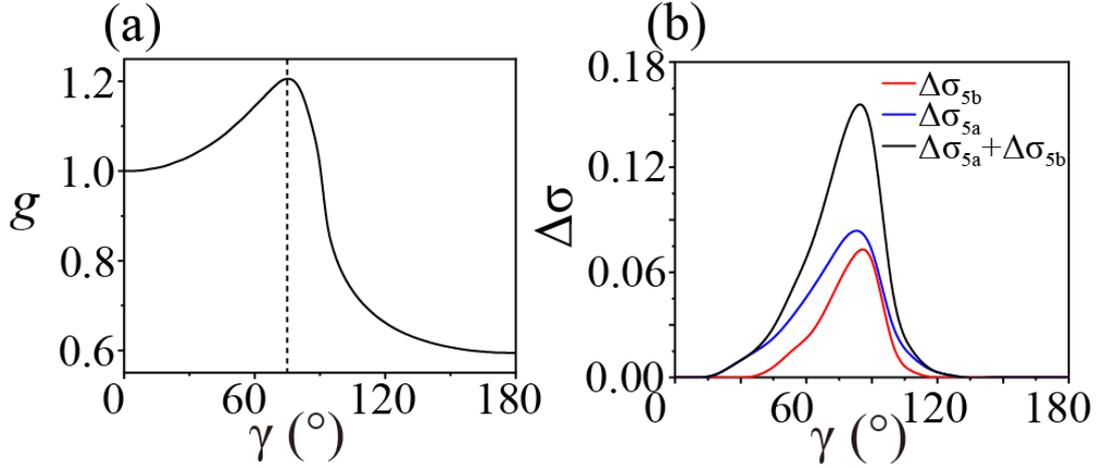

Fig. 1S (a) compression coefficient $g$ of the fifth band of the moiré superlattice with (324.8°, 86.4°) as a function of $\gamma$. (b) the red line represents the difference between the concentration factors of the two split modes at the edge of the Brillouin zone below the fifth band: $\Delta\sigma_{5b} = \sigma_{5bb} - \sigma_{5ba}$. The blue line represents the difference between the concentration factors of the two split modes at the edge of the Brillouin zone above the fifth band: $\Delta\sigma_{5a} = \sigma_{5ab} - \sigma_{5aa}$. The black line represents $\Delta\sigma_{5b} + \Delta\sigma_{5a}$.

**Section 2. Experimental samples and their transmission spectra on the TPB**

We choose tantalum pentoxide as material a and silicon dioxide as material b, that is, $n_a = 2.17$, $n_b = 1.46$. We fabricated three superlattices samples: sample A (212.8°, 111.6°), sample B (295.2°, 93.6°), sample C (324.8°, 86.4°), which are marked in Fig. 1(i). For samples, the sample A has 28 periods and the $\Lambda = 332nm$, and the sample B has 14 periods, and the $\Lambda = 737nm$, and the sample C has 8 periods, and the $\Lambda = 1379nm$. The red shaded areas in Fig.2S(a)-(c) are the transmission spectra on the TPBs of different samples. Since the bands are compressed by α, the bandwidth of the transmission spectrum becomes narrower as α approaches 360°. In Fig. 2S(a)-(c), the solid black line is the experimental measurement data, and the dashed line is the calculation result of the transfer matrix.

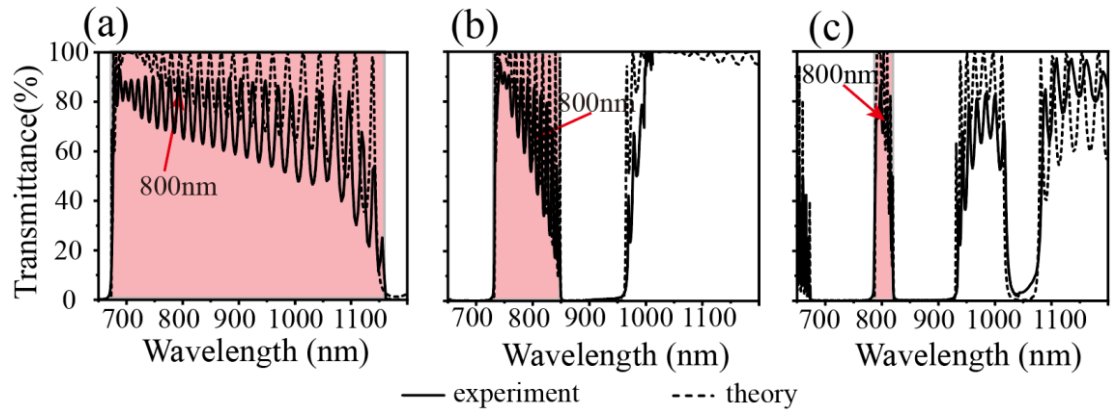

Fig. 2S (a)-(c) The transmission spectrums on the TPB of sample A, sample B, sample C, respectively. The solid line is the experimental measurement data, and the dotted line is the theoretical calculation results.

We now consider the effect of a TPB with an oblique incidence for different polarizations. Taking the moiré superlattice with two synthetic angles (α, γ): (324.8°, 86.4°) as an example, Fig. 3S(a) is the oblique incidence band for the TE mode and Fig. 3S(b) is the oblique incidence band for the TM mode. The green area is the TPB in Fig. 3S (a, b). Obviously, for TE mode, as the oblique incidence angle $\theta$ increases, the TPB is narrower. Their transmission spectra corresponding to different angles $\theta$ are shown in Fig. 3S(c, d). The line is the experimental measurement, and the dashed line is the theoretical calculation. For the TE mode, the transmission spectrum narrows with angle $\theta$ increasing and for the TM mode is the opposite.

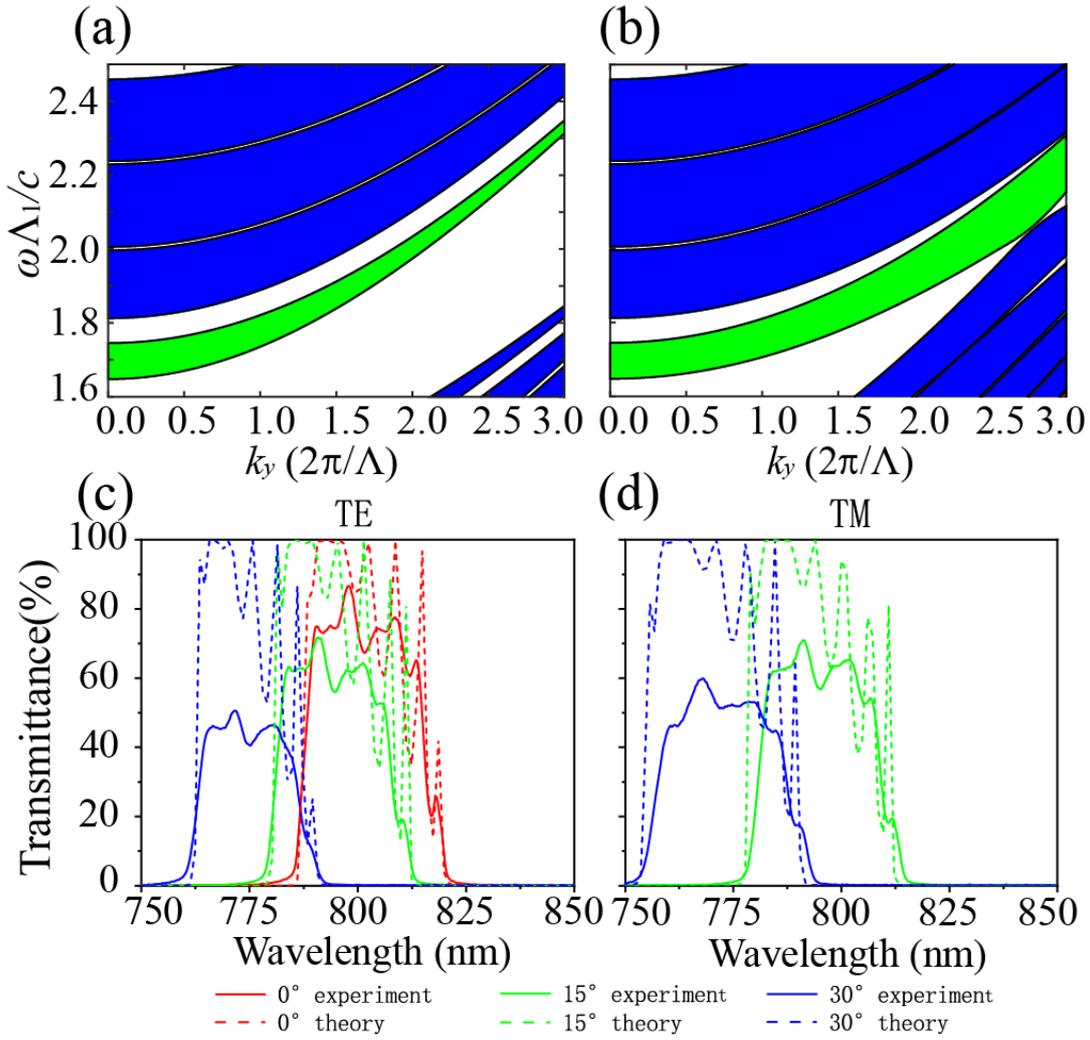

Fig. 3S (a) For TE mode, band of oblique incidence of the moiré superlattice with two synthetic angles (α, γ): (324.8°,86.4°) for TE mode. (b) For TM mode, band of oblique incidence of the moiré superlattice with (324.8°,86.4°). (c) For TE mode, the transmission spectrum of oblique incidence on the TPB of the moiré superlattice with (324.8°,86.4°). (d) For TM mode, the transmission spectrum of oblique incidence on the TPB of the moiré superlattice with (324.8°,86.4°).

**Section 3. The effect of oblique incidence on the group index on the TPB**

Considering the effect of oblique incidence on the group index on the TPB, we calculated the group index for different angles with two synthetic angles (α, γ): (324.8°,86.4°), as shown in Fig. 4S(a). The two polarizations show different trends. As shown in Fig. 4S(b), for TE mode, the group index on the TPB increases with the oblique incidence angle $\theta$, the faster it increases with the angle of α. This is because for TE mode, the TPB narrows with increasing oblique incidence angle $\theta$, the slope of the TPB decreases, and the group velocity decreases. However, for TM mode, the TPB widens with increasing incidence angle $\theta$, the slope of the TPB increases, and the group velocity increases

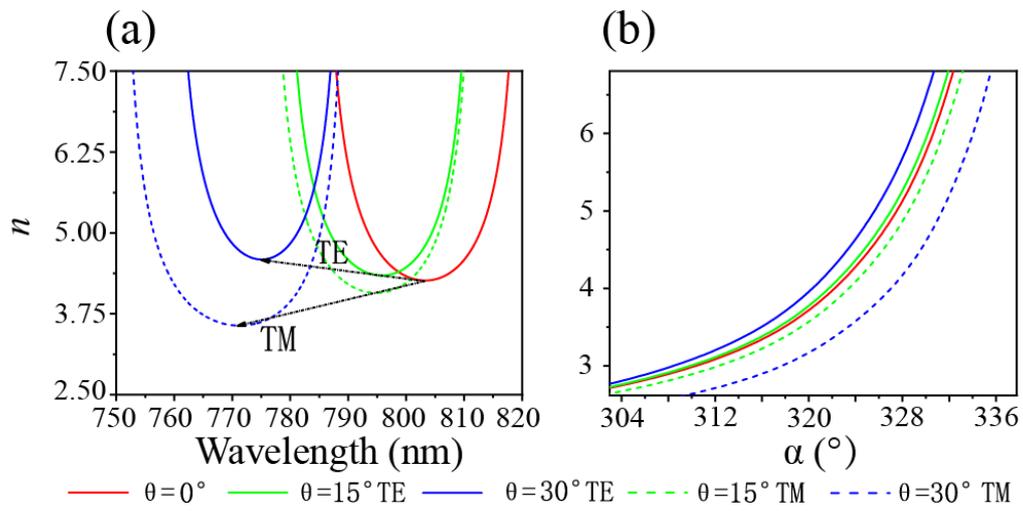

Fig. 4S(a) Group indices with different oblique incidence angles on the TPB of the moiré superlattice with two synthetic angles ($\alpha$, $\gamma$): (324.8°,86.4°), the two polarizations, TE modes and TM modes show different trends. (b) Group index on the TPB for different oblique incidence angles $\theta$ increase with $\alpha$. The solid line indicates TE mode. The dotted line indicates TM mode.